\documentstyle[12pt,aasms4]{article}
\tighten

\newcommand\be{\begin{equation}}
\newcommand\ee{\end{equation}}
\newcommand\jcd{Christensen-Dalsgaard}

\input epsf
\lefthead{Antia and Basu}
\righthead{Temporal variations of the solar rotation rate}
\begin{document}

\title{Temporal variations of the rotation rate in the solar interior}

\author{H. M. Antia} 
\affil{Tata Institute of Fundamental Research, 
Homi Bhabha Road, Mumbai 400005, India}
\and
\author{Sarbani Basu}
\affil{Astronomy Department, Yale University, P.O. Box 208101 New Haven\\
CT 06520-8101, U. S. A.}

\begin{abstract}
The temporal variations of the rotation rate in the solar interior 
are studied using frequency
splittings from Global Oscillations Network Group (GONG) data obtained
during the period 1995--99. We find alternating latitudinal bands of 
faster and slower rotation
which appear to move towards the equator with time --- similar to
the torsional oscillations seen at the solar surface. This  flow
pattern appears to persist to a depth of about $0.1R_\odot$ and in
this region its magnitude is well  correlated with
solar activity indices. We do not find any periodic or systematic
changes in the rotation rate near the base of the convection zone.
\end{abstract}


\section{Introduction}

Helioseismic data allow us to probe the rotation rate in the solar interior
as a function of radius and latitude (Thompson et al.~1996; Schou et al.~1998).
With the accumulation of Global Oscillations Network Group (GONG) and
Michelson Doppler Imager (MDI) data over the last five years, it 
has also  become
possible to study the temporal variation in the rotation rate and other
properties of the  solar interior.

It is well known that frequencies of solar oscillations vary with time
(Elsworth et al.~1990; Libbrecht \& Woodard 1990; Dziembowski et
al.~1998; Bhatnagar, Jain \& Tripathy 1999; Howe, Komm \& Hill 1999).
These variations are also known to be  correlated with solar activity.
However, these frequency changes appear to result from variations in
solar structure close to the surface;
there is no clear evidence for structural changes in deep interior
(Basu \& Antia 2000a).
Inversions for the rotation rate in solar interior on the other hand, have
confirmed that the temporal variations in the rotation
rate  penetrate to somewhat  deeper layers
(Howe et al.~2000a; Toomre et al.~2000). Furthermore, the changing pattern
of solar rotation rate in interior also agrees with the torsional
oscillations observed at the solar surface (Howard \& LaBonte 1980).
While most of these studies are restricted to the outer part of
the convection zone, the seat of the solar dynamo is generally believed to
be near the base of the convection zone and rotation is believed to
play an important role in the operation of the solar dynamo.
This is also the region where the tachocline is located (Kosovichev 1996;
Basu 1997).
Thus one needs to
look for possible changes in rotation rate in this region.
Recently, Howe et al.~(2000b) have reported a 1.3 year periodicity
in variation of equatorial rotation rate at a radial distance of $0.72R_\odot$.
It is not clear if the period of 1.3 years is associated with solar
cycle variations. Considering the fact that we have only 4 years of
data, this period is also uncomfortably  close to the orbital 
period of
the Earth and one may expect systematic errors on this period being
introduced in the data. If this periodicity is confirmed, it may have 
implications for the theories of solar dynamo. Hence, this needs to be
checked by an independent analysis.

In this work we use oscillation frequencies obtained from GONG 
observations  covering a period from May 1995 to
June 1999, to study possible temporal variations in the rotation rate
as a function of radius and latitude. We use both 1.5d and
2d inversion using the Regularized Least Squares (RLS) technique
(Antia, Basu \& Chitre 1998) to determine the rotation rate from the
observed splitting coefficients.

\section{The data and technique}

We use data for GONG months 1--44 to determine the rotation rate in the solar
interior. Each of these data sets covers a period of 108 days and is
centered on
dates 36 days apart. Thus there is significant overlap between neighboring
data sets. Each GONG month covers a period of 36 days with month 1, starting
on May 7, 1995 and month 44 ending on September 6, 1999. We have used 42
data sets centered on GONG months 2--43 to study the temporal variations
in the rotation rate.  The GONG project
determines the frequencies of each mode $\nu_{n\ell m}$ (Hill et al.~1996)
where $n,\ell,m$ are respectively, the radial order, degree and azimuthal
order of the mode. The frequencies of all observed modes in an
$n,\ell$ multiplet are fitted to an expansion in terms of orthogonal
polynomials (Ritzwoller \& Lavely 1991) in $m$, 
\begin{equation}
\nu_{n\ell m}
= \nu_{n\ell} + \sum_{j=1}^{j_{\rm max}} c_j (n,\ell) \,
{\cal P}_j^{(\ell)}(m),
\label{eq:acof}
\end{equation}
where $c_j$ are the splitting coefficients which are determined by
least squares fit to the frequencies $\nu_{n\ell m}$, and the number of
terms $j_{\rm max}$ is set such that
$j_{\rm max}\le2\ell$.  For this work, we have fitted splitting coefficients
up to $c_{16}$; beyond this there is very little signal in the fitted
coefficients,
and because of the orthogonality of these polynomials, inclusion of higher
coefficients does not make any difference to the fitted values of the
lower order coefficients.
The odd-order coefficients $c_1,c_3,\ldots$ are determined
by the rotation rate in solar interior and hence these can be used 
to determine rotation rate as a function of latitude and radius.
The splitting coefficients are sensitive only to the north-south symmetric
component of rotation rate and hence that is the only component that can
be determined through inversions. We therefore
restrict ourselves to this symmetric component.
The anti-symmetric component in the outer layers can be studied using other
techniques
such as ring diagram analysis (e.g., Basu, Antia \& Tripathy 1999;
Basu \& Antia 2000b).
We use splitting coefficients for modes with $1.5<\nu<3.5$ mHz and
$\ell\le150$ only to perform the inversions of the different GONG
data sets.

Traditionally (e.g., Snodgrass 1984), the latitudinal dependence
of the solar surface rotation rate
is expressed as a second degree polynomial
in $\cos^2\theta$, $\theta$ being the colatitude. This is
the ``smooth'' part of the rotation rate. The torsional
oscillations at the solar surface have been studied using the residuals
left after subtracting the mean rotation rate obtained from this second
degree polynomial fit (Howard \& LaBonte 1980).
Since the signal in these residuals is rather weak other prescriptions
for calculating the ``mean'' rotation rate have also been tried to
study the sensitivity of resulting pattern to these changes
(LaBonte \& Howard 1982; Snodgrass 1992).
The polynomial approximation to the rotation rate
is essentially equivalent
to using splitting coefficients $c_1,c_3$, and $c_5$ only for inversions.
Thus we can determine the smooth part of rotation rate by using only
these three coefficients and take a mean over the period for which
data are available. Since  helioseismic data are available only for
a small fraction of the magnetic cycle, the average will not represent a true
mean of rotation rate over long time intervals, but that is the best we
can do with the available data sets.  This mean can be subtracted from the 
rotation rate at any given
epoch to get the time varying component of the rotation rate. Thus we
can write
\begin{equation}
\delta\Omega(r,\theta,t)=\Omega(r,\theta,t)-
\langle \Omega_s(r,\theta,t)\rangle,
\label{eq:zonal}
\end{equation}
where $\Omega$ is the rotation rate at any given epoch and $\Omega_s$ is
the smooth part of rotation rate as a function of latitude. Here the angular
brackets denote average over the time duration for which data are available.
This time varying component $\delta\Omega$ is generally called the zonal flow.
We will refer to this time varying component variously  as the residual 
in rotation rate or the zonal flow. It may be noted that this residual
cannot be directly compared with the observations at solar surface
as different mean rotation would have been subtracted in the two cases.

We use both  1.5d and 2d RLS techniques
as described by Antia et al.~(1998) to invert the splitting
coefficients to determine the rotation rate as a function
of radius and latitude for each data set.  The  advantage of the
1.5d inversion is that
each coefficient is treated separately and hence we can combine the
results from required coefficients as per the need for smooth part
and full rotation rate. In  the 2d inversion technique one has to do separate
inversions to calculate these parts independently. As a result,
for 2d inversion it is more  convenient to take the temporal mean
of the entire rotation rate and subtract it from the rotation rate at
each epoch to get the residual velocity. This prescription
has also been used in literature (e.g., LaBonte \& Howard 1982;
Howe et al.~2000a)
to isolate temporal variation in rotation rate. Unless otherwise
stated, all results in this paper have been obtained by subtracting
the mean of only the  smooth component of rotation rate.
The use of two independent
inversion techniques enable us to test the results for possible
systematic errors in inversion.

Observations from the Earth or the first Lagrange point, where the 
SOHO satellite is located, will only give the
synodic rotation rate, which needs to be converted to sidereal value.
This conversion may introduce some error, which will have periodic
variations on the time-scale of one year.
There may be other systematic errors in
inverted rotation rate, but most of these may be independent of time
and may get cancelled when the mean rotation rate is subtracted.

\section{Results}

\begin{figure}
\plotone{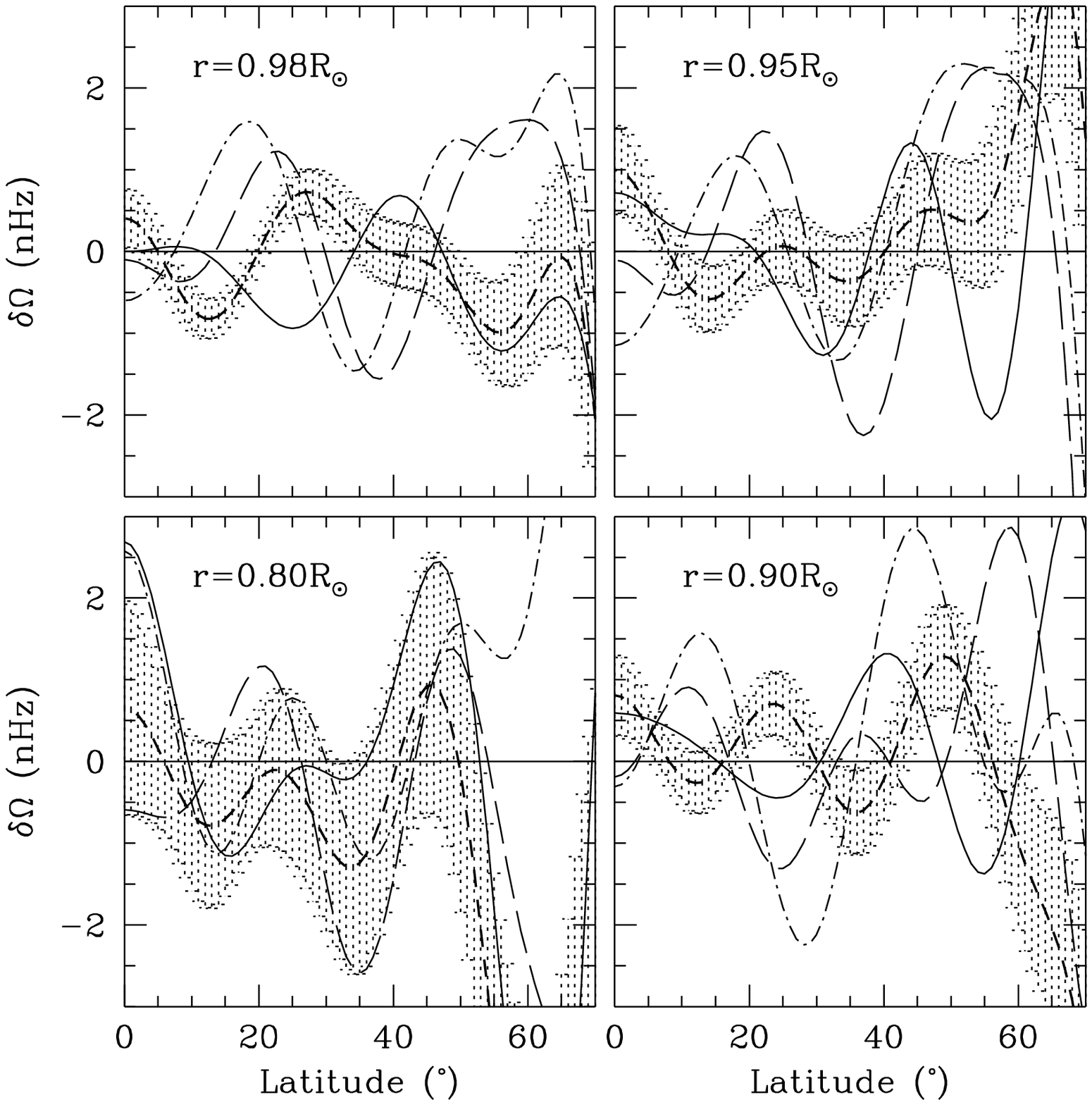}
\figcaption{
The residuals in the rotation rate (zonal flow)
shown as a function of latitude at a few
selected depths and time.  The continuous curves are for GONG months 1--3
centered around  June 1995, short-dashed for June 1996,
long-dashed for June 1997 and dot-dashed for June 1998. Error bars
are shown only for the short-dashed curve for the sake of
clarity. The error-bars are similar for all cases.
\label{fig:zonalcutbw}
}
\end{figure}

The zonal flows obtained by using the 1.5d inversion
technique are shown in Fig.~\ref{fig:zonalcutbw} as a
function of latitude for a few data sets and depths. 
These results were obtained by  subtracting the average of the
smooth rotation rate (as obtained from coefficients $c_1$, $c_3$ and
$c_5$) from the the full rotation rate (eq.~\ref{eq:zonal}).
It is clear that the residuals
are significant and the time-variations 
in the layers immediately below the surface are well
above the noise level.
It also appears that the magnitude of $\delta\Omega$ increases with
solar activity and the peaks in the latitudinal profile tend to 
shift towards
the equator with time.

\begin{figure}
\plotone{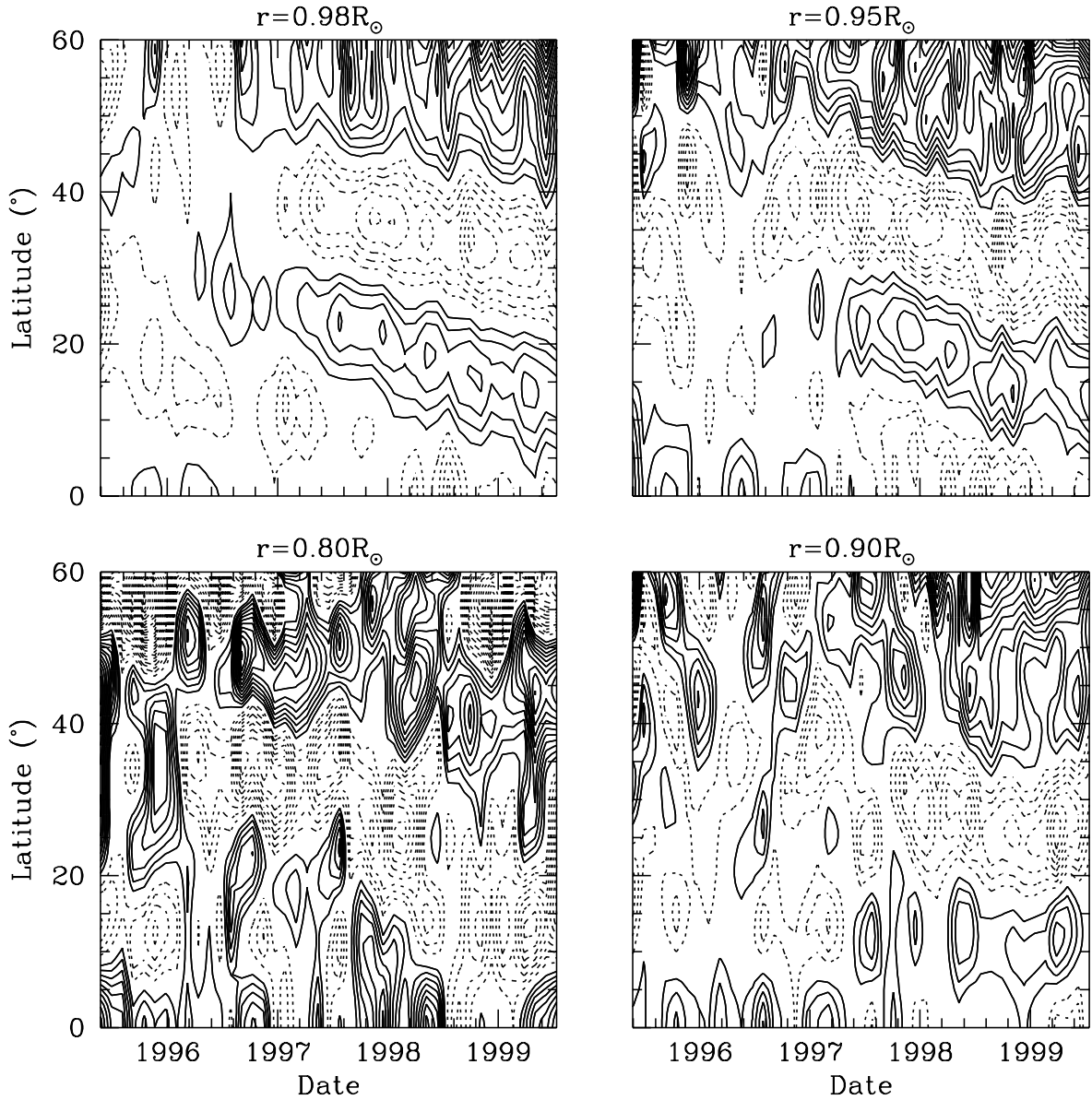}
\figcaption{
Contours of constant $\delta\Omega$  shown as a
function of time and latitude at a few selected radius as marked above
each panel.  The continuous contours are for positive values
and dotted for negative values.
The contours are drawn at interval of 0.4 nHz.
\label{fig:zonalbw}
}
\end{figure}

The zonal flow pattern is more clearly seen in Fig.~\ref{fig:zonalbw}, 
which shows the
contours of constant $\delta\Omega$ at a few selected depths
as a function of latitude and time.
At any given time the flow pattern has alternating bands in latitude where the
rotation rate is either faster or slower than the average. These bands
appear to move towards the equator, and at some time after the solar
activity minimum the equatorial region changes from faster
to slower rotation. This pattern is similar to the well known
torsional oscillations observed at the solar surface (Howard \& LaBonte 1980).
A similar pattern
has been seen in other global helioseismic analyses too (Schou 1999;
Howe et al.~2000a; Toomre et al.~2000), and also in the near surface layers 
studied using 
local-helioseismology methods (Basu \& Antia 2000b). There is 
a good agreement
between different independent results obtained in the outer layers of
the Sun.

\begin{figure}
\plotone{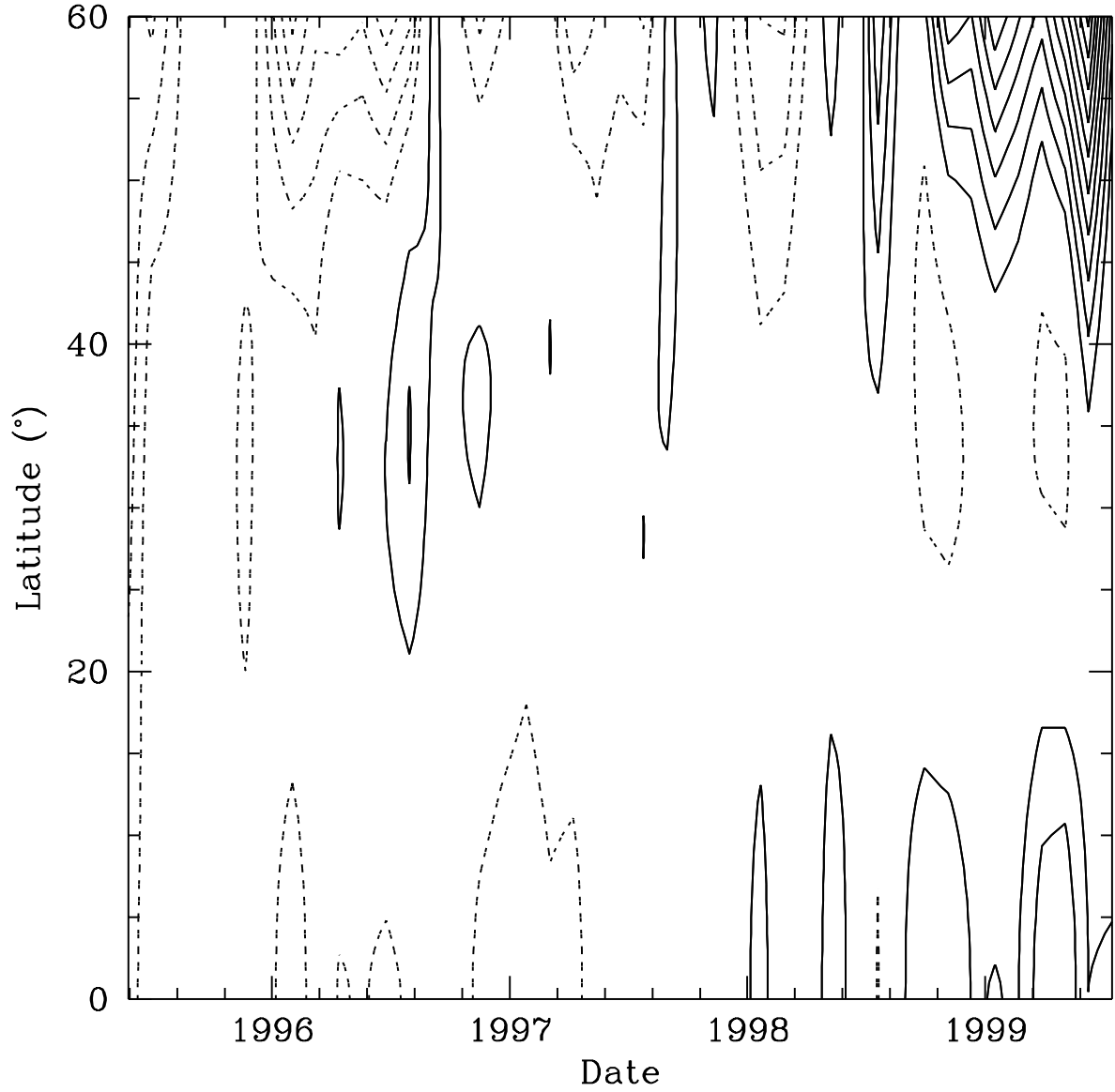}
\figcaption{
A contour diagram of the residuals in the smooth component
of the rotation rate.
The results are shown as a function of time
and latitude for a radius of $r=0.98R_\odot$.
The continuous contours are for positive values
and dotted for negative values.
The contours are drawn at interval of 0.4 nHz.
\label{fig:zonala1to5}
}
\end{figure}

Unfortunately, the prescription for isolating the time-dependent
component from
the full rotation rate is not unique and in many cases the mean of 
the entire rotation rate (not just the smooth part) is subtracted to 
obtain the residual.
A problem with this prescription is that the mean depends on the time
duration over which the averaging is done, unless the data covers a
very long period. On the other hand, the smooth component of rotation
rate shows much less temporal variation and hence the mean of the smooth
part is relatively insensitive to duration of data.
The variation of the smooth component of the flow with
time can be seen in Fig.~\ref{fig:zonala1to5}.
This figure shows the residuals obtained by subtracting the time-averaged
smooth component from the smooth component at each epoch.
The time variation in this case is much smaller
than the variation in the full rotation rate. For example, at low
latitudes the full rotation rate varies by about $\pm2$ nHz,
while the smooth component varies by only $\pm0.5$ nHz. This clearly shows
that most of the temporal variation is in the higher-order
splitting coefficients. However, the temporal variation seen in
Fig.~\ref{fig:zonala1to5} is still larger than the error estimates
and hence is statistically significant. Of course, the resulting
pattern is different from that seen in Fig.~2 with full rotation profile
as smooth component by definition cannot yield more than two zeros in
each hemisphere. There will also be significant distortion due to the
fact that the mean over a short period of 4 years is subtracted
to get the residuals. The variation seen in Fig.~\ref{fig:zonala1to5}
is consistent with that seen in smooth component of surface rotation
rate (LaBonte \& Howard 1982) in the latitude range shown in Fig.~3.

\begin{figure}
\plotone{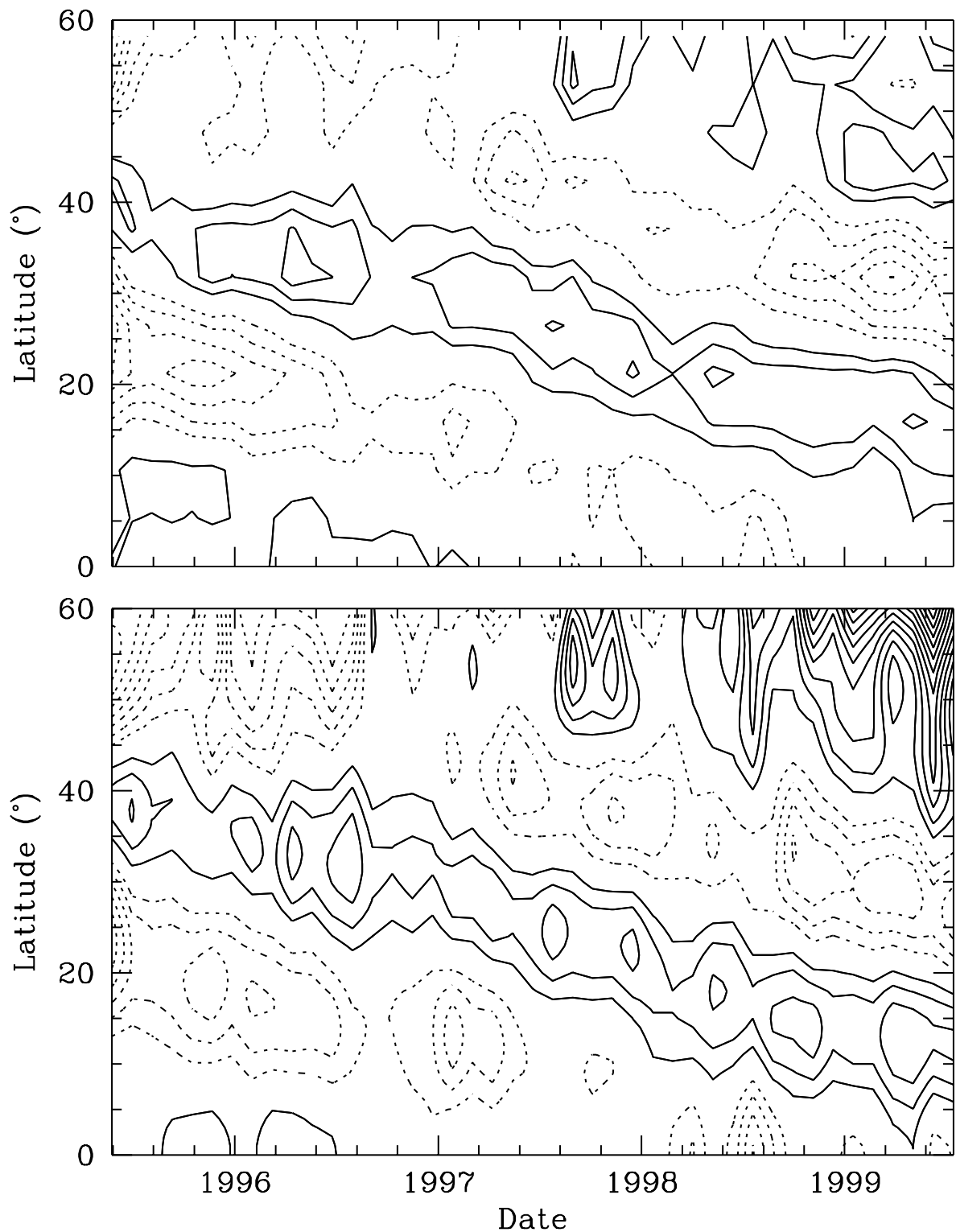}
\figcaption{
A contour diagram of the residuals  obtained
by subtracting the time-average of the rotation rate from the rotation
rate at a given time. The upper panel is the result obtained
by 2d RLS inversion, while the lower panel is from 1.5d RLS inversion.
The results are for radius $0.98R_\odot$.
The contours have been drawn at intervals of 0.4 nHz, with continuous
lines for positive contours and dotted lines for negative ones.
\label{fig:zonalcomp}
}
\end{figure}

\begin{figure}
\plotone{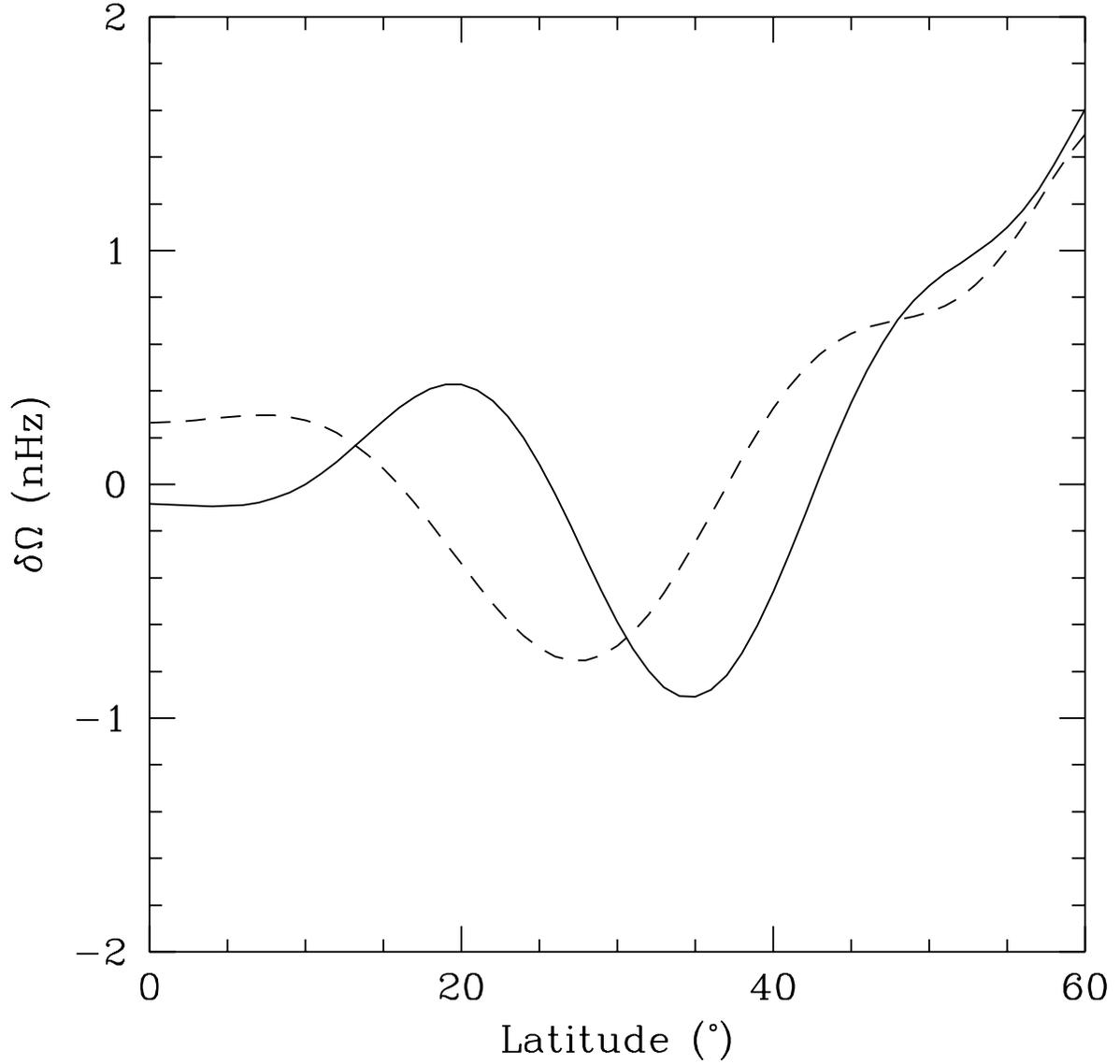}
\figcaption{The difference in temporal averages of rotation rate
obtained from the full profile and the smooth component
($\langle\Omega(r,\theta,t)\rangle - \langle\Omega_s(r,\theta,t)\rangle$)
at two different depths are shown as a function of latitude.
The continuous and dashed lines respectively, represent the difference
at $r=0.98R_\odot$ and $0.90R_\odot$.
\label{fig:rotavg}
}
\end{figure}

In order to test the sensitivity of results to varying prescriptions
for isolating
the time-dependent component, we also compute the residuals where the
mean of the full rotation rate is subtracted and the results at
$r=0.98R_\odot$ are shown in Fig.~\ref{fig:zonalcomp}.
These results can  be compared with
the upper left hand panel in Fig.~\ref{fig:zonalbw}.
It can be seen that although the
general pattern is similar in both cases, there are significant differences
in the details. In Fig.~\ref{fig:zonalbw}, which shows the results
obtained by subtracting only the mean of the
smooth component of rotation, the amplitude of residual
appears to increase with activity, whereas this increase is not seen in
the pattern when the mean of the full rotation rate is subtracted.
In order to understand the difference between the two prescriptions for
calculating zonal flow component of rotation rate, we show in
Fig.~\ref{fig:rotavg} the difference in temporal averages of the full
rotation rate and
that of the smooth component at a few selected depths. This difference
will have to be added/subtracted to convert from one definition to the 
other.
It is clear that since the difference itself shows latitudinal variations
with amplitude comparable to the zonal flow pattern, there will be a
significant difference between the residuals when different averages
are subtracted.
Fig.~\ref{fig:zonalcomp} also
compares the results at $r=0.98R_\odot$ obtained using different
inversion techniques. 
We find that the zonal flow pattern obtained using 2d RLS inversions 
agree reasonably well with those obtained with 1.5d RLS, though there are
differences at high latitudes. Figs.~8,9 also compare the 1.5d and 2d
RLS results and once again it is clear that there is reasonable
agreement between the two sets of results.
Apart from this we have also varied the regularization used in each
inversion technique to check the sensitivity of results to choice of
smoothing. It is found that results are not particularly 
sensitive to the choice of smoothing used in the inversions. Similarly,
the results are also insensitive to the choice of modes (frequency range)
used for the inversions.

\begin{figure}
\plotone{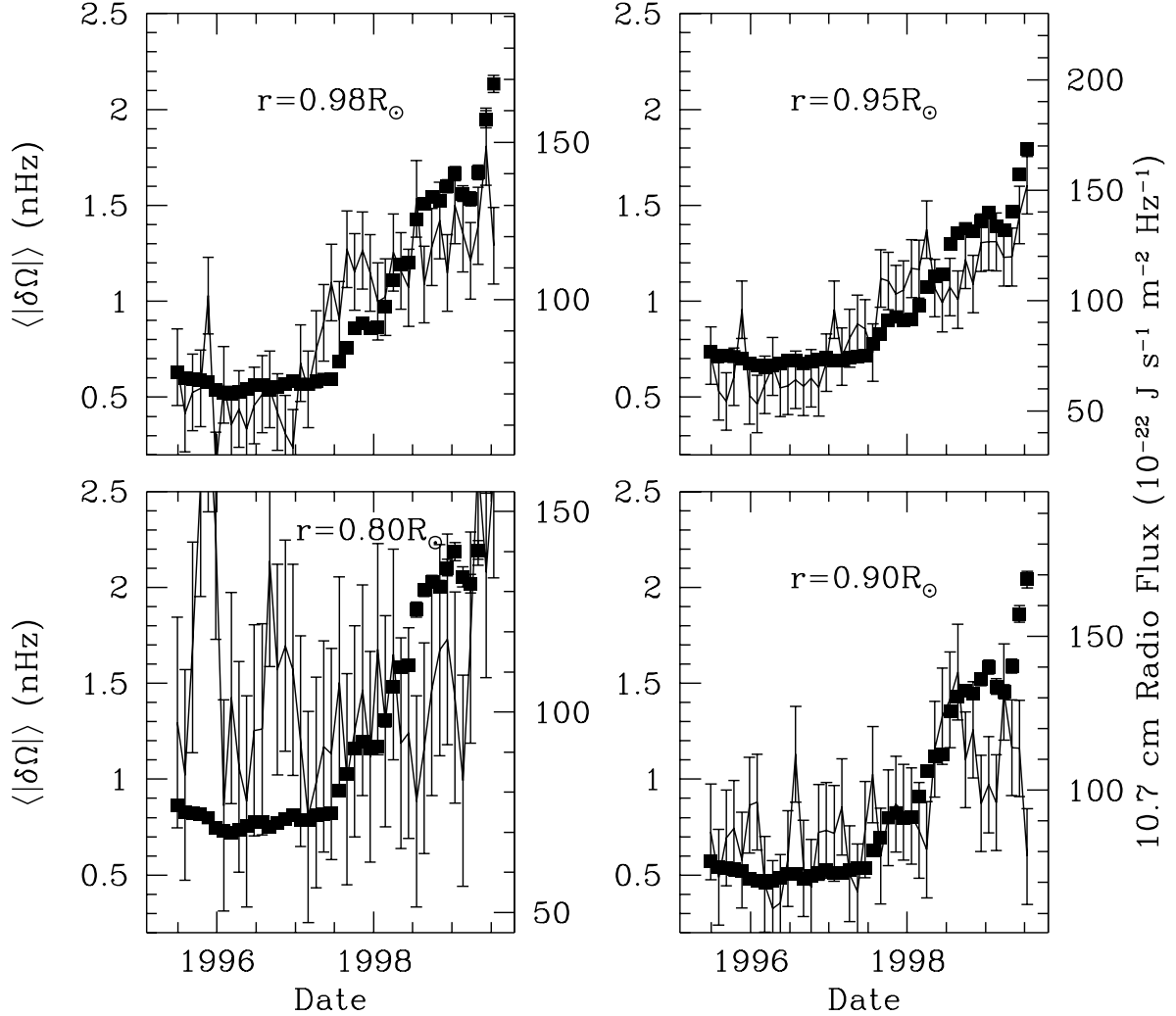}
\figcaption{
The mean magnitude of the residual in rotation rate (in the latitude range of
0--60$^\circ$) at different depths plotted
as a function of time  and compared with an indicator of solar activity.
The  points joined by lines with error-bars show the zonal flow results
for each of the available data sets.  The squares show 
the average 10.7 cm flux during the corresponding period
on a scale shown on the right side.
Each panel corresponds to the radius marked inside the panel. 
\label{fig:zonalssnbw}
}
\end{figure}

As mentioned earlier, from Figs.~\ref{fig:zonalcutbw} and
\ref{fig:zonalbw}, we find that the amplitude of the zonal flow 
increases with solar activity. 
In fact, the flow pattern is almost absent near the
solar activity minimum.  In order to study this variation,
in Fig.~\ref{fig:zonalssnbw} we show the average
magnitude of this component over the latitude range of
0--60$^\circ$ as a function of time. This figure also shows the
mean radio flux at 10.7 cm during the time interval covered by each data set
as obtained from
Solar Geophysical data web page (www.ngdc.noaa.gov/stp/stp.html) of the
US National Geophysical Data Center. The 10.7 cm flux from the
Sun is known to track solar activity.
It is clear that there is a reasonable
correlation between the mean magnitude of the zonal flow
in layers immediately below the solar surface and the 10.7 cm flux.
A similar correlation is found if we use the maximum magnitude of zonal flow
velocity instead of the mean magnitude, or when other solar activity
indices like the daily sunspot number are used instead of the 10.7 cm
flux. This correlation is not particularly sensitive to the range 
of latitude used in averaging. Further, the magnitude of mean zonal
flow velocity decreases with depth.
Similar correlation between the mean magnitude of zonal flow
velocity and solar activity has been seen in results obtained using
ring diagram technique (Basu \& Antia 2000b).

\begin{figure}
\plotone{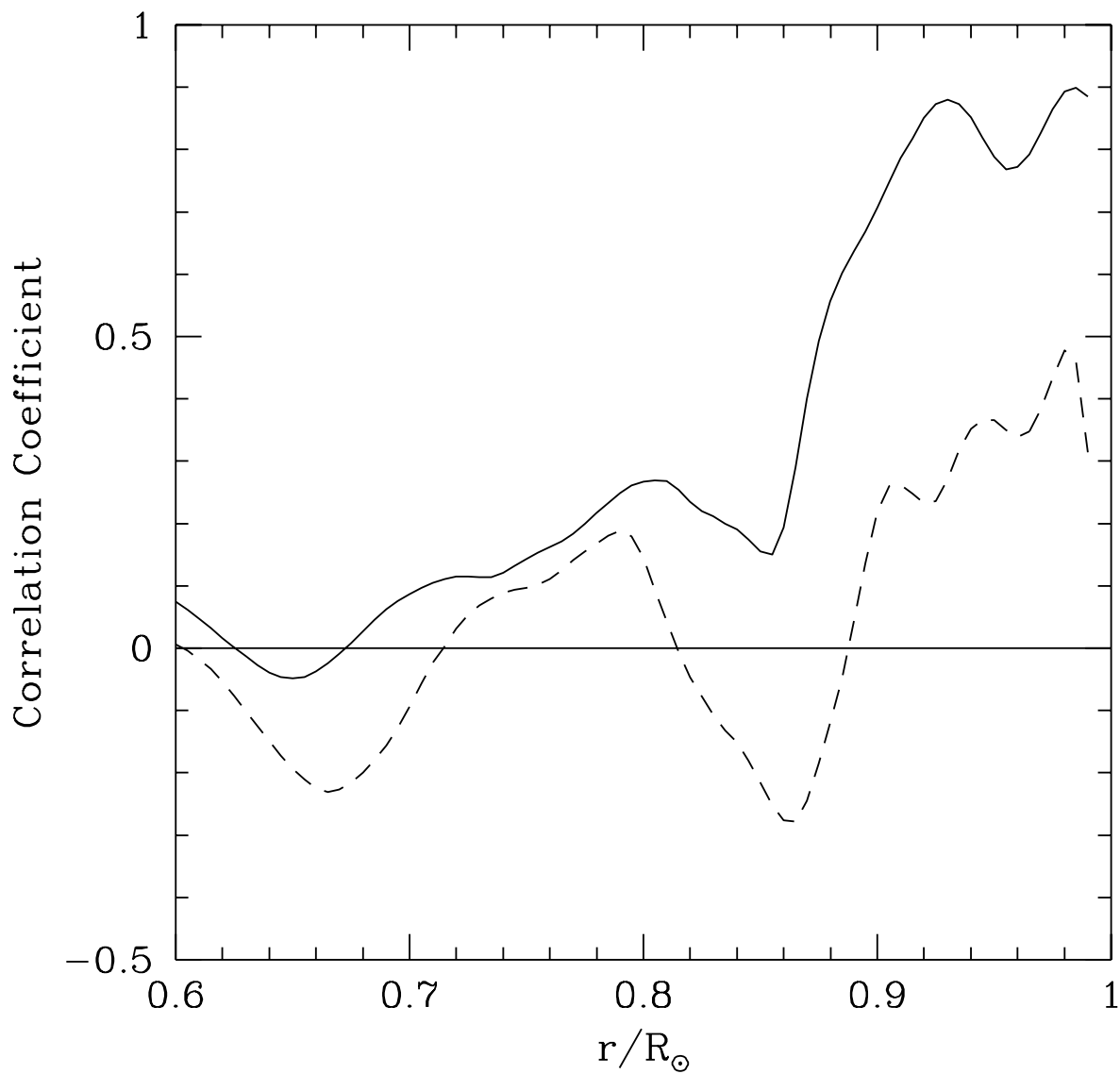}
\figcaption{
The correlation coefficient between the mean magnitude of the residuals
in rotation rate in the
latitude range 0--60$^\circ$ and the average 10.7 cm radio flux during
the period the data were obtained 
is plotted as a function of radius. The continuous line represents the result
obtained when the residual is determined by subtracting the mean of
only the smooth component of the rotation rate (obtained from coefficients 
$c_1,c_3,c_5$) from the total rotation rate. The 
dashed line is the result when the time-average of the full rotation
rate is subtracted from the rotation rate at a given time.
\label{fig:zonalcor}
}
\end{figure}

At depths exceeding $0.1R_\odot$ the correlation between the zonal flow
and the 10.7 cm flux becomes markedly weaker and there appears
to be no correlation at a depth of $0.2R_\odot$. To show this result
quantitatively, in Fig.~\ref{fig:zonalcor} we plot the correlation coefficient
between the average  10.7 cm flux over the period
observations were made and the latitudinally averaged 
magnitude of $\delta\Omega$ as a
function of radius. This figure shows that the correlation
between the zonal flow and solar activity does not extend to the
bottom of the convection zone. It probably also implies that
the flow pattern itself  does not penetrate through the
convection zone. Similar conclusions can be drawn from Fig.~\ref{fig:zonalbw},
where the flow pattern  is seen to change significantly below a radius of
$0.9R_\odot$. The correlation coefficient shows a sudden drop around
the radius of $0.89R_\odot$ and this could be considered to be the depth
to which flow pattern penetrates.
There is some difficulty in estimating the penetration-depth of the
zonal flows to high precision because the errors
in inversion increase with depth and it becomes increasingly difficult to
detect any pattern by just looking at the results at different
depths. But the sudden drop in the  correlation coefficient
around $0.89R_\odot$ is difficult to explain as being
caused by an increase in the errors.  Neither the errors, nor the resolution
drops suddenly below $0.89R_\odot$ and hence we would expect a
slower drop due to these factors.
We can thus conclude that the zonal flow pattern penetrates
to a depth of about $0.1R_\odot$.

As mentioned earlier, if instead of subtracting the mean rotation rate
obtained from only the coefficients $c_1,c_3,c_5$, we subtract the mean
of full rotation rate, the amplitude of residuals doesn't show any
increase with activity. This can also be seen in Fig.~\ref{fig:zonalcor}
which also shows the correlation coefficient for this case.
This is due to the fact that the higher order
coefficients $c_7,c_9,\ldots$ show an increase in amplitude with
activity.

\begin{figure}
\plotone{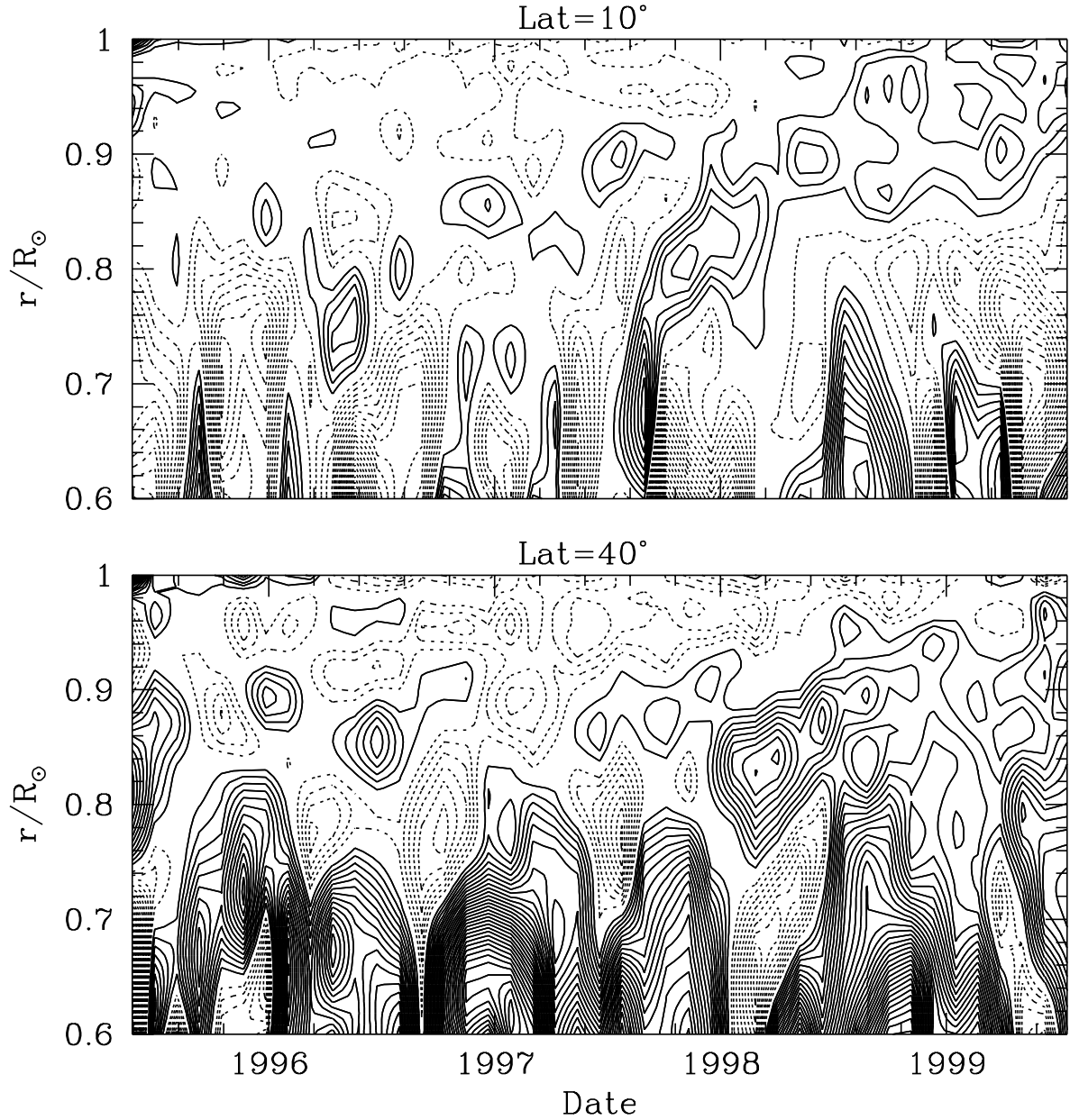}
\figcaption{
The residual in rotation rate plotted as a 
function of time and radius at a few selected latitudes as marked
above each panel.  The continuous contours are for positive
values and dotted contours for negative values.
The contours are drawn at intervals of 0.4 nHz
\label{fig:zonalrbw}
}
\end{figure}

In order to study the radial behavior of the zonal flow,
we have plotted the flow in Fig.~\ref{fig:zonalrbw}  as a
function of radius and time at a
few selected latitudes. It can be seen that the pattern is quite different
below about $r=0.9R_\odot$. 
This is just another indication that the 
flow pattern persists only in the outer layers
of the Sun. At all latitudes there appears
to be a tendency of the bands of faster or slower rotation to move
up in radius with time. Though the significance of this upward movement
is not clear as it is not seen at all times.
Around the tachocline and below that there appear
to be some pronounced oscillations with time at all latitudes. It is
not clear if these are real or arise due to the regularization used
in the inversion.

The oscillations in the region around the tachocline have also been 
studied  by
Howe et al.~(2000b) who claim to see a 1.3 year periodicity in the equatorial
rotation rate at a depth of $0.72R_\odot$. In order to verify this
periodicity, in  Figs.~\ref{fig:zonalcuta} and \ref{fig:zonalcutb} we show
the residuals in the rotation rate at a few selected depths
and latitudes as a function of time. For clarity the error-bars are
shown only for the 1.5d RLS inversions. The estimated errors in 2d
inversions are comparable to those in 1.5d RLS at $r=0.95R_\odot$ and
$0.90R_\odot$, while in deeper layers (Fig.~\ref{fig:zonalcutb}) the
error-bars on the results of the 2d inversions are about a factor of 
2 lower than those of the 1.5d inversions.
In order to make a proper comparison with Howe et al.~(2000b) these
residuals have been obtained by subtracting the time average of the 
full rotation rate from the rotation rate obtained from
each data set.
It is quite clear that there is a
systematic variation with time in the rotation rate of the  outer layers (which
is clearly seen in Fig.~\ref{fig:zonalbw} also though that was
obtained by subtracting just the smooth part of the mean rotation
rate). In deeper 
layers there are some variations which do not seem to be  systematic. 
In order to test whether there are any periodic variations,
we take a discrete Fourier transform of the data for the
equator at $r=0.72R_\odot$. The resulting amplitude is
shown in Fig.~\ref{fig:zonaldft}.
There is no prominent peak in this spectrum at a  period of 1.3 yr, or at 
any other period.
Similar spectrum is obtained for other latitudes and depths in this
region.  Some of the apparent
variations seen in the region around the tachocline can arise from the
fact that in this region the rotation rate changes very rapidly with radius.
Inversions tend to yield some average of the rotation rate
in the neighborhood, and 
furthermore, since each data set has slightly different error estimates and
mode set, the averaging would be different and one can expect to get
some spurious variations.

\section{Conclusions}

\begin{figure}
\plotone{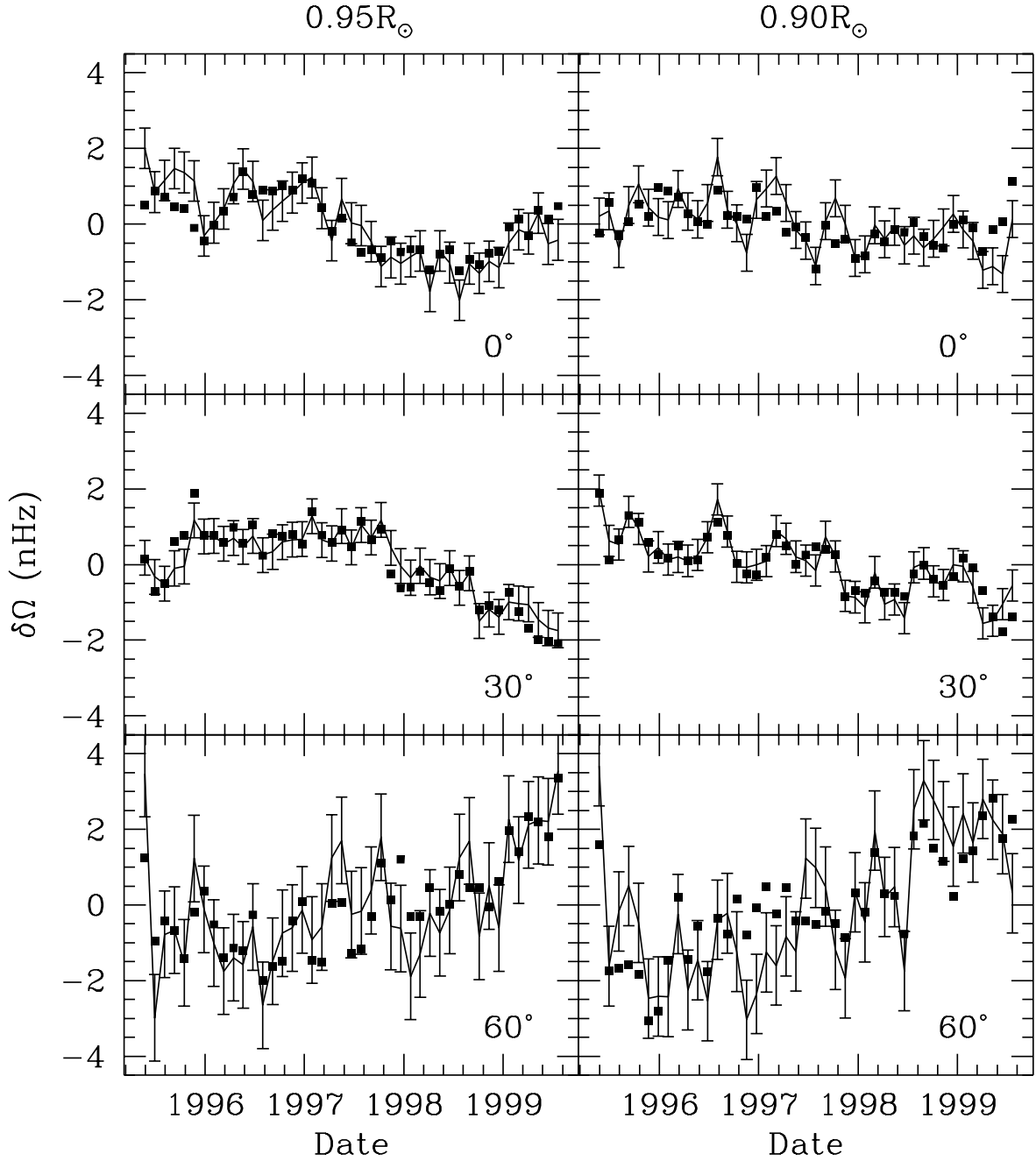}
\figcaption{
The residuals in the rotation rate shown as a function of time at a 
few selected
depths and latitudes.  These residuals were obtained by subtracting the
time-average of the full rotation rate from the rotation rate obtained
from each data set.
The continuous lines with error-bars show the results
obtained using 1.5d RLS inversion, while the squares mark the results from
2d RLS inversion. The latitude is marked in each panel, while the radius
is marked just above the top panel.
\label{fig:zonalcuta}
}
\end{figure}

\begin{figure}
\plotone{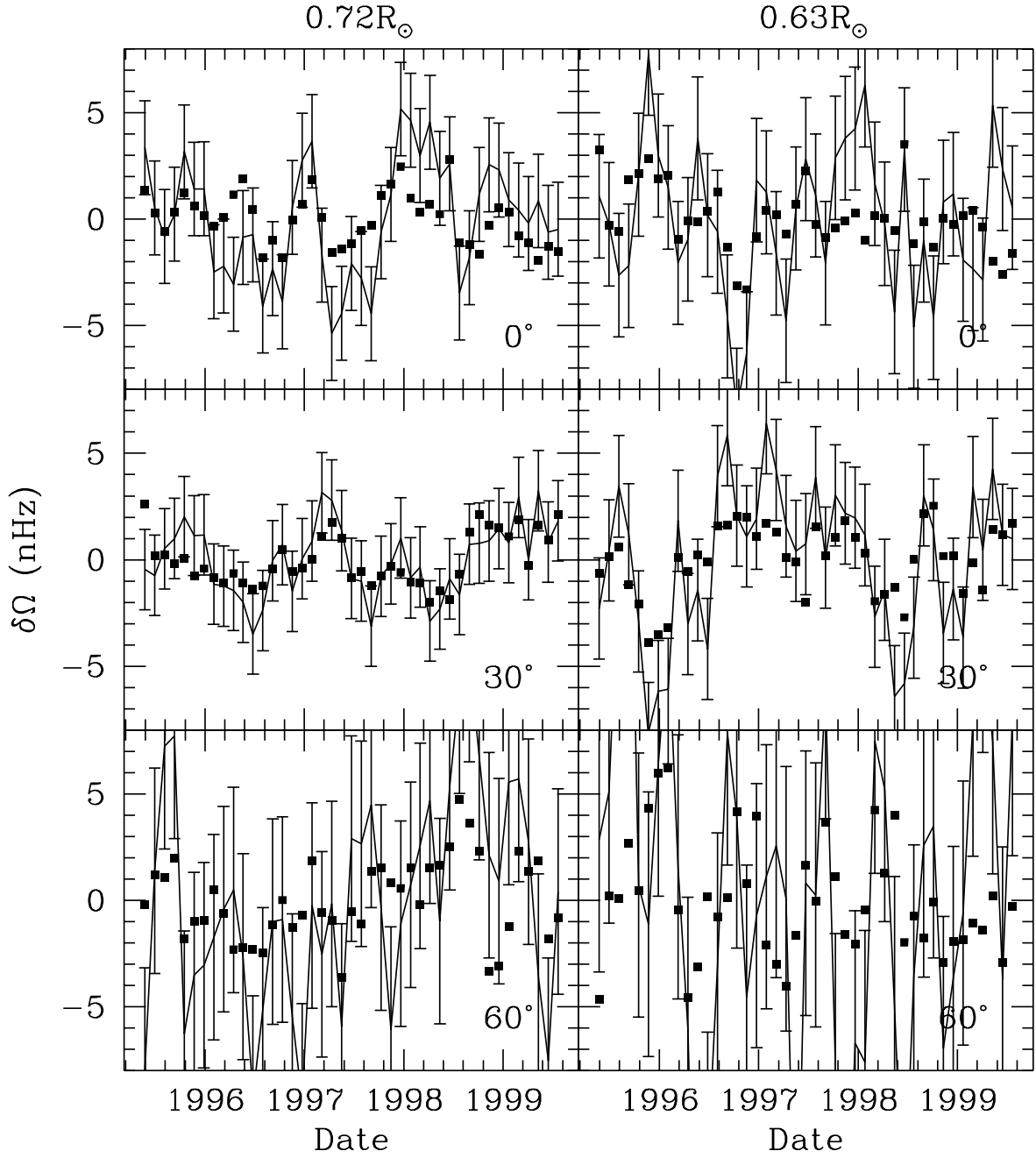}
\figcaption{The same as Fig.~(\ref{fig:zonalcuta}) but for
different depths.
\label{fig:zonalcutb}
}
\end{figure}

\begin{figure}
\plotone{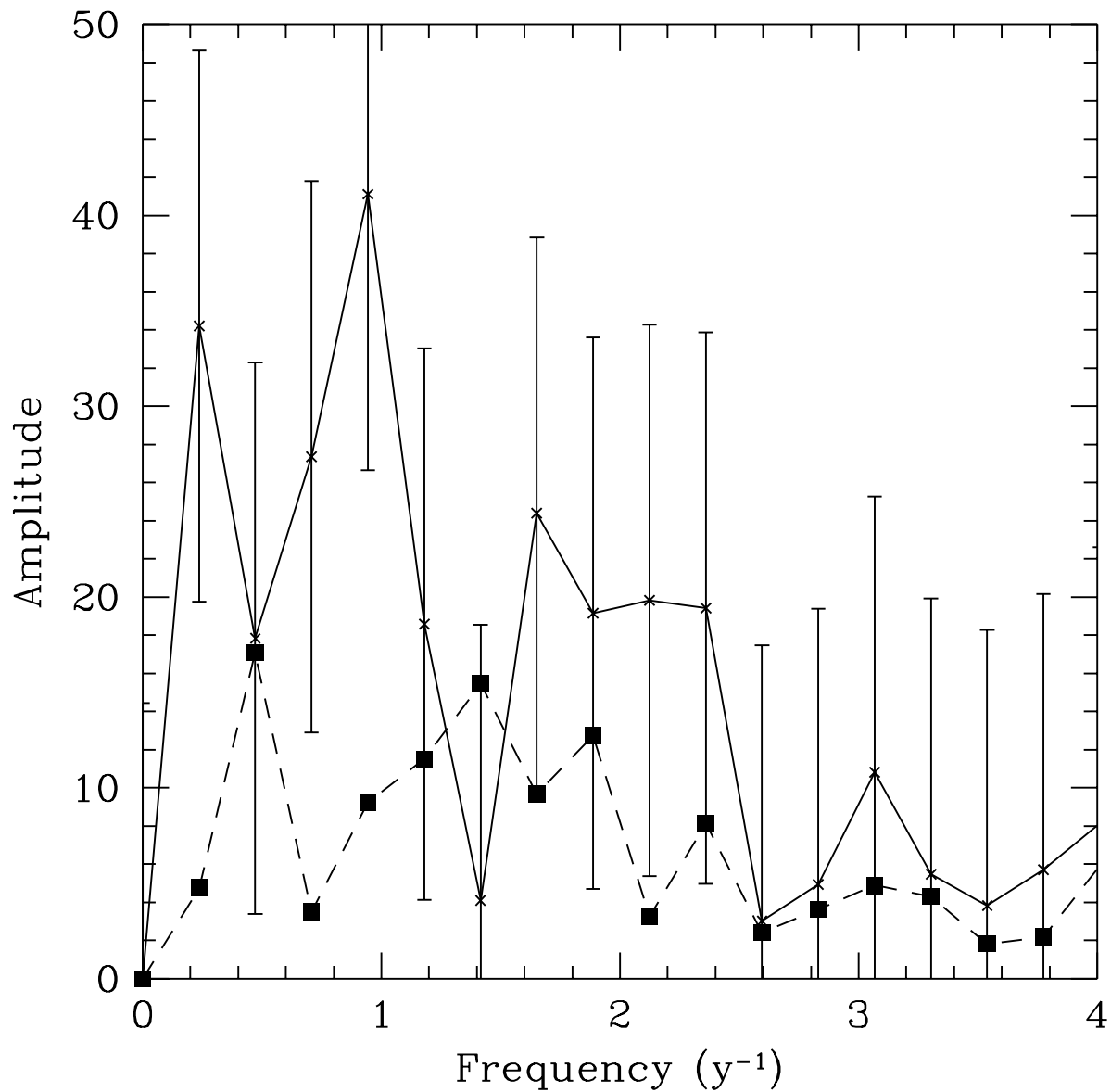}
\figcaption{Results of a discrete Fourier transform of the 
rotation-rate residuals
at the equator at $r=0.72R_\odot$. The continuous line shows the result
using 1.5d RLS inversion while the squares mark the results from 2d
RLS inversion. For clarity the error-bars are shown only for 1.5d results.
\label{fig:zonaldft}
}
\end{figure}

Inversions to determine the solar rotation rate with data from GONG
months 1--42, which mostly cover the rising phase of
solar cycle 23,  show a clear change in the rotation rate  with solar activity.
The zonal flow, which we have defined as residual in the rotation rate 
obtained by subtracting a time-average of a three-term fit
to the rotation rate, shows bands of faster and slower rotation. 
These bands appear to move towards the equator with time.
The magnitude of this time varying component is about 2 nHz.
This pattern is similar to that found by Howe et al.~(2000c) using
helioseismic data from GONG and MDI projects. The amplitude of the flow
pattern in our results is also similar to those in Howe et al.~(2000c)
as can be seen by comparing their Fig.~3 with our Fig.~\ref{fig:zonalcuta}.

The mean magnitude of the pattern in the outer parts of the solar
convection zone
appears to increase with solar activity and is reasonably well
correlated with the mean 10.7 cm flux which tracks solar activity.
The correlation is
seen  up to a depth of about $0.1R_\odot$ (70 Mm).
This appears to be depth to which the zonal flow pattern penetrates.
Below this depth the temporal variations in the rotation rate do not appear
to be significant. This inferred depth of zonal flow pattern is consistent
with estimate of at least 60 Mm obtained by Howe et al.~(2000c) using
independent data sets and inversion techniques. Clearly, there is 
good agreement between different inversion results in the outer
convection zone.
There appears to be  some hint that the flow pattern moves
upwards with time, however, the statistical significance of this
is as yet unclear.

We do not find any evidence of periodic changes in the solar
rotation rate near the base of the solar convection zone. 
We find that the temporal variations 
at this depth are random in nature and probably represent statistical
fluctuations. There is no clear periodic signal in the residual rotation rate
with period less than 2 years at any depth. 
The present data length
of 4 years does not allow us to study periodicities with longer time
period. Some of this fluctuations may also arise from systematic
errors in data.

\acknowledgments

This work utilizes data obtained by the Global Oscillation Network
Group (GONG) project, managed by the National Solar Observatory, a
Division of the National Optical Astronomy Observatories, which is
operated by AURA, Inc. under a cooperative agreement with the National
Science Foundation. The data were acquired by instruments operated by
the Big Bear Solar Observatory, High Altitude Observatory, Learmonth
Solar Observatory, Udaipur Solar Observatory, Instituto de Astrofisico
de Canarias, and Cerro Tololo Interamerican Observatory.

\end{document}